# von Neumann first outlined the possible non existence of dispersion free ensembles in quantum mechanics: may we verify non existing dispersion free ensembles by application of quantum mechanics in experiments at perceptive and cognitive level?

*Elio Conte* [1,2], *Orlando Todarello* [2], *Antonio Federici* [2],
*Nunzia Santacroce* [1], *Vincenza Laterza* [1]
and
*Andrei Yuri Khrennikov* [3]

[1] School of Advanced International Studies for Applied Theoretical and non Linear Methodologies of Physics, Bari – Italy
[2] Department of Neurological and Psychiatric Sciences. University of Bari – Italy.
[3] School of Mathematics and Systems Engineering, University of Vaxjo, S-35195 Sweden.

**Abstract:**
von Neumann in 1932 was the first to outline the possible non existence of dispersion free ensembles in quantum mechanics, and he used also such basic evidence to give a preliminary proof on incompatibility between quantum mechanics and (non-contextual) hidden variables theory. In the present paper we give a detailed theoretical elaboration on the manner in which such fundamental subject could be explored at perceptive and cognitive level in humans. We also discuss a general design of the experiment that we have in progress so to give direct indications to other researchers engaged in such field.

**Key words**: application of quantum mechanics in psychology and neuroscience, dispersion free ensembles, quantum cognition, quantum contextuality, language analysis by quantum mechanics.

## Introduction.

It is well known that in quantum mechanics Kochen-Specker (KS) theorem (Kochen, Specker 1967) and A. Peres (1991, 2002) and N.D. Mermin (1990) results rule out the *non-contextual* assignment of values to physical observables. One cannot assume, without falling into contradictions, that an observed entity enjoy a separate well-defined identity irrespective of any particular context observing it.
We have classical against quantum intrinsic contextuality.
In classical contextuality, the outcome is affected by various aspects of the environment, but not by the irreducible and nonpredictable specifics of the interaction. between the system and the experimental observation. When the situation is analyzed in terms of states, experiments, and outcomes, Kolmogorov's axioms are satisfied, and a classical probability model can be used. Started with 1999 D. Aerts (Aerts et al 2000, 2005, 2011) and A. Khrennikov (Khrennikov et al. 1999, 2002, 2004, 2006, 2010) attempted to introduce such basic notion of contextuality and thus quantum mechanics at the level of human cognitive performance. Further studies in this direction, also recalling what previously formulated by us in 1983 were elaborated by us at the theoretical and experimental level (Conte et al 1983,2000, 2003, 2004, 2006, 2007, 2008, 2009, 2010, 2011), confirming the central role that could be explained from quantum mechanics in cognitive performance of humans.
Let us explain shortly the basic differences between quantum and classical contextuality.
As an example of classical contextuality in a cognitive situation, we may consider a task in which we ask to a subject if he likes a shown object. His/her answer would depend much more on the object than on how the question will be posed. Of course we may conceive other situations in which the outcome is determined through the interaction of the system with the irreducible and nonpredictable properties embedded in the measurement process. This is a case in which we are concerned instead with as intrinsic contextuality. The system and its observation both have an internal relational constitution, such that their interface creates a concrescence of emergent, dynamic patterns. The presence of intrinsic contextuality means that Kolmogorovian axioms are not satisfied, which renders the formal description of the entity nonclassical, but quantum mechanical. This is a very important conclusion that of course recalls a result that we have recently obtained

*There are stages of our reality in which we no more can separate the logic ( and thus cognition and thus conceptual entity) from the features of "matter per se". In quantum mechanics the logic, and thus the cognition and thus the conceptual entity-cognitive performance, assume the same importance as the features of what is being described. We are at levels of reality in which the truths of logical statements about dynamic variables become dynamic variables themselves so that a profound link is established from its starting in this theory between physics and conceptual entities*.

The human beings are adept and drawing context-sensitive and cognitive associations.

## Theoretical Elaboration

Consider now a given quantum system $S$. Physically speaking, the Hilbert space of the system $S$ then contains wave functions belonging to different possible contexts. We may conceive such states as lying in a given energy shell in phase space. Let the considered dimensionality to be ($m$). Each possible state of the system is then represented by a unit vector in $\Lambda_0(m)$. The wave function $\psi(t)$ of the system lies in the appropriate space $\Lambda_0(m)$ having dimension ($m$). Suppose the orthonormal vectors $\omega_1, \omega_2, \ldots, \omega_m$ are a basis for $\Lambda_0(m)$. The projection of $\psi(t)$ in a co-ordinate axis $\omega_j$ is

$$c_{t,j}(a) \equiv (\psi(t), \omega_j) ; \quad j = 1, 2, \ldots, m \tag{1}$$

and it depends directly on the considered context (a). The probability of finding the system at time $t$ will be given by

$$p(t) = \sum_{j=1}^{m} |c_{t,j}|^2 \tag{2}$$

and depends on the context ($a$).

Consider now all the possible contexts. If such all contexts exist at all in principle, there will be many of them, and thus we may average over the different contexts. Consequently one may calculate the average probability of finding the system at a certain time, and it will be given by

$$<p(t)> = \sum_{j=1}^{m} <|c_{t,j}|^2> \tag{3}$$

The spread of the individual values of $p(t)$ due to the different contexts will be given by the second moment

$$Z = \frac{<(p(t) - <p(t)>)^2>}{<p(t)>^2} = \frac{<p(t)^2>}{<p(t)>^2} - 1 \tag{4}$$

and it may be estimated experimentally.

In detail, consider also the notion of dispersion free ensembles. According to von Neumann

Given the observable $\Re$,

An ensemble is dispersion free if

$$<\Re^2> = <\Re>^2 \tag{5}$$

Reasoning in analogy, we conclude that $Z \to 0$ in this case.

Let us take now a little step on.

Being the present only a brief technical note, we will not enter here in the specialized details of the very specialized field of mental lexicon. It represents a well advanced field of research and applications that we cannot consider here for brevity. We will assume that it is well known to the reader and, in any case, we quote here the excellent papers of some authors (Bruza & Cole 2005), suggesting the reader to read such papers and obviously the whole outfit of experimental and theoretical research that has been previously developed and quoted in detail in such papers.

We limit to outline here that language is inherently contextual. According to the previous mentioned authors, consider the word 'bat'. This word has at least two senses in its standard form. It might refer to a flying mammal that lives in caves, or alternatively it might refer to a sporting implement. This is only a very restricted example but actually we may relate the word "bat" to a very large number of contexts. Generally we can tell the sense that another speaker intends through a consideration of the context in which the word appears. These different *senses* of a word can be explored via word association experiments (Bruza & Cole 2005).

In free association, words are presented to large samples of participants who produce the first associated word to come to mind. The probability or strength of a pre-existing link between words is computed by dividing the production frequency of a response word by its sample size We can also find out which words are likely to produce the word 'bat' (now called a target). One way of achieving this involves a process known as *extra-list cuing*. Here, subjects typically study a list of to-be-recalled target words. For widening on the experiments, we still indicate to read carefully ref.7. and references therein.

## Design of the Experiment and Preliminary Conclusions

Starting with a general model based upon the notion of superposition of states as it is intended in quantum mechanics, the authors in (Bruza & Cole 2005) introduced a model of the observed behaviour of associative networks and they also developed more sophisticated models of how concepts combine (Bruza & Cole 2005). These models treat context by representing it as cue words, or coappearing words, and experiments go currently underway to test their validity
The basic assumption is that the words take different senses depending upon the context in which they occur. As example, when shown out of context, 'bat' reminds people of 'cave', sporting people and so on. The problem is to estimate the probability of recalling 'bat' when some context is present. With reference to the initial elaboration given in (1)-(4) let us restrict our "semantic" Hilbert space to $m = 2$, and identify by $\omega_1$ the cognitive state of recalling and by $\omega_2$ the complementary state of not recalling.
The recall (or not) of a word can be represented using a superposition state,

$$\psi(t) = c_{1,t}\omega_1 + c_{2,t}\omega_2 \tag{6}$$

with specific notations given in the (2).
This is the word $w$, represented in some context $(a)$, as a superposition of recalled, and not recalled. Thus, the word 'bat' is a target word, expected to be recalled in an extra-list cueing experiment upon presentation of the cue word 'cave' which in this case acts as the context $(a_1)$. We have

$$\psi(t) = c_{1,t}(a_1)\omega_1 + c_{2,t}(a_1)\omega_2 \tag{7}$$

The probability of 'bat' being recalled in this context is represented by $|c_{1,t}(a_1)|^2$, and the probability of not being recalled by $|c_{2,t}(a_1)|^2$,
When given the cue word 'ball' we represent 'bat' as
the new superposition

$$\psi(t) = c_{1,t}(a_2)\omega_1 + c_{2,t}(a_2)\omega_2 \tag{8}$$

where $(a_2)$ represents the new context "ball" and the new probabilities result now modified as $|c_{1,t}(a_2)|^2$, and $|c_{2,t}(a_2)|^2$, respectively, and assuming obviously totally different values respect to the previous case just as they may be retrieved from memory when a subject is presented with the cue 'ball' than the cue word 'cave'.
We would suggest here a general formalism to represent that it provides a very natural representation of contextual effects as they actually occur in language.
It is evident that we could continue with the word "bat", considering each time a different context ($a_1, a_2, \ldots, a_n$), and thus obtaining each time a different representation of the assumed quantum superposition (recalling-not recalling) with different values of the coefficients $c_{1,t}(a_i)$ and $c_{2,t}(a_i)$ ($i = 1, 2, \ldots, n$), and each time different values of probabilities. Consequently, we may apply the (3) in the case $m = 2$, considering that we may estimate finally the average probability, valued on all the considered contexts. As consequence, selected a given context ($a_i$), we may calculate the spread of the individual value by using the coefficient $Z$ as indicated in (3) by the second moment of the considered experimentation.
The most simple assumption is that we have a very large number of alternative possibilities, and that, in order to properly characterize such situation, we must use a continuous their distribution. Without loss of generality we may consider

$c_1 = \cos\alpha$ and $c_2 = sen\alpha$; $p_{recalling} = \cos^2\alpha$; $p_{not-recalling} = sen^2\alpha$; $p_{+,-}$ being probabilities

where any one value of $\alpha$ ($0 \leq \alpha < 2\pi$) now characterizes a different, possible context. The probability of finding an angle in the range ($\alpha, \alpha + \delta\alpha$) will be given by

$$f(\alpha) = A sen^b(2\alpha)d\alpha \tag{9}$$

where $A$ is a normalizing constant and for a strictly uniform distribution we have ($b = 0$), while for a weakly uniform distribution we have possible values ($b = 2$, or $b = 4$) and so on. Under the different theoretical as well as experimental situations, we may also consider more restricted range of possible values for $\alpha$ as ($0 \leq \alpha < \pi$) or ($0 \leq \alpha < \pi/2$), and so on.

First consider the very interesting case in which the possible contexts obey a law of strictly uniform distribution. Generally speaking, we know that, given the density function of probability $f(x)$, it must be

$$\int_a^b f(x)dx = 1 \quad , \quad <x> = \int_a^b x\,f(x)dx \quad , \quad <x^2> = \int_a^b x^2 f(x)dx \tag{10}$$

In the case of strictly uniform distribution ($b = 0$), we obtain that for ($0 \leq \alpha < 2\pi$)

$$A = 1/2\pi \quad , \quad <p_+> = <p_-> = 1/2 \; ; \; <p_+^2> = <p_-^2> = 3/8 \tag{11}$$

In this case, under experimentation, we expect to be:
  a) The (5) is violated (not existing dispersion free ensembles)
  b) The $Z$-value, given in (4) furnishes $Z = 0.5$
  c) Finally, $<(p_{+,-}(generic\ \alpha) - 0.5)^2> = 1/8$.

Let us examine now the case of a weakly uniform distribution in ($0 \leq \alpha < 2\pi$), ($b = 2$) in the (9). It results that

$$A = 1/\pi \quad , \quad <p_+> = <p_-> = 1/2 \; ; \; <p_+^2> = <p_-^2> = 5/16$$

Under experimentation, we expect to be:
  d) The (5) is violated (not existing dispersion free ensembles)
  e) The $Z$-value, given in (4) furnishes $Z = 0.25$
  f) Finally, $<(p_{+,-}(generic\ \alpha) - 0.5)^2> = 1/16$.

For $b = 4$, we have that

$$A = 4/3\pi \quad , \quad <p_+> = <p_-> = 1/2 \; ; \; <p_+^2> = <p_-^2> = 7/24$$

Under experimentation we should find that
  g) The (5) is violated (not existing dispersion free ensembles)
  h) The $Z$-value, given in (4) furnishes $Z = 0.16$
  i) Finally, $<(p_{+,-}(generic\ \alpha) - 0.5)^2> = 0.04$.

We do not expect the results to change radically exploring contexts in the range ($0 \leq \alpha < \pi/2$)

Finally let us evidence further the conceptual foundations and the malleability of the formulation that we have introduced.

Let us examine this time our elaboration for contexts ranging with (($0 \leq \alpha < \pi/4$), ($b = 2$). It results that

$$A = 8/\pi \quad , \quad <p_+> = 0.71; \; <p_-> = 0.29; \; <p_+^2> = 0.52; <p_-^2> = 0.10$$

Under experimentation we should find now that

  j) The (5) is violated (not existing dispersion free ensembles)
  k) $Z_+ = 0.03 \quad\quad ; \quad Z_- = 0.19$
  l) $<(p_+(generic\ \alpha) - 0.71)^2> = 0.0151 \; ; \; <(p_-(generic\ \alpha) - 0.29)^2> = 0.0159$

It is clearly seen that the situation is now profoundly modified.

Obviously, we could have selected contexts ranging instead with ($\pi/4 \leq \alpha < \pi/2$) as well as, in principle, we could experience also a different kind of analytical expression of density probability function instead of the (9).

The conclusion is that, starting with the basic formulation given by the authors in (Bruza & Cole 2005), we have given here a little conceptual extension that may be well built by arranging appropriate experiments.

The most promising evidence is that by this methodology we may also analyze the presence or not of dispersion free ensembles. We remember here a datum that may be of basic interest when exploring quantum cognition. von Neumann in 1932 (1996) was the first to outline the possible non existence of dispersion free ensembles in quantum mechanics, and he used also such basic evidence to give a preliminary proof on incompatibility between quantum mechanics and (non-contextual) hidden variables theory.

# References

Aerts D., D'Hondt E., & Gabora L. Why the disjunction in quantum logic is not classical. Foundations of Physics 2000; 30 (9): 1473-1480. [quant-ph/0007041]

Aerts D. & Gabora L. A state-context-property model of concepts and their combinations I: The structure of the sets of contexts and properties. Kybernetes 2005; 34 (1&2): 167-191. (Special issue dedicated to Heinz Von Foerster.) [quant-ph/0402207] [pdf]

Aerts D. & Gabora L. A state-context-property model of concepts and their combinations II: A Hilbert space representation. Kybernetes 2005; 34 (1&2): 192-221. (Special issue dedicated to Heinz Von Foerster.) [quant-ph/0402205] [pdf]

Aerts D., Broekaert J. & Gabora L. A case for applying an abstracted quantum formalism to cognition. New Ideas in Psychology 2011; 29 (1).

Bruza P.D. & Cole R.J. Quantum logic of semantic space: An exploratory investigation of context effects in practical reasoning In S. Artemov, H. Barringer, A. S. d'Avila Garcez, L.C. Lamb, J. Woods (eds.) We Will Show Them: Essays in Honour of Dov Gabbay. College Publications 2005.

Bruza, P., Kitto, K., Nelson, D., & McEvoy, C. Is there something quantum-like in the human mental lexicon? Journal of Mathematical Psychology 2009; 53: 362-377.

Bruza P., Widdows D., Woods J. A quantum logic of down below. In: Engesser K., Gabbay D., Lehmann D., editors. Handbook of Quantum Logic and Quantum Structures 2009; 2: 625-660. arXiv:quant-ph/0612051.

Bruza P.D., Busemeyer J., Gabora L. Journal of Mathematical Psychology 2009; 53 Special issue on quantum cognition.

Conte E., exploration of the biological function by quantum mechanics using biquaternions. Proceedings of the 10[th] International Congress on Cybernetics, entitled Le concept d'organisation in Cybernetique, edited by the Association International de Cybernetique. Namur-Belgique, August 22-27[th], 1983; 16-24.

Conte E. Biquaternion quantum mechanics. Pitagora Editrice, Bologna,-Italy, 2000.

Conte E, Todarello O, Federici A, Vitiello F, Lopane M, Khrennikov AY. A Preliminar Evidence of Quantum Like Behaviour in Measurements of Mental States, Quantum Theory, Reconsideration of Foundations. Vaxjio Univ. Press, 2003; 679-702.

Conte E., Vena A., Federici A., Giuliani R., Zbilut J.P. A brief note on a possible detection of physiological singularities in respiratory dynamics by recurrence quantification analysis. Chaos, Solitons and Fractals 2004; 21 (4): 869-877

Conte E., Federici A., Zbilut J.P. On a simple case of possible non-deterministic chaotic behavior in compartment theory of biological observables. Chaos, Solitons and Fractals 2004; 22: 277-284.

Conte E., Pierri GP., Federici A., Mendolicchio L., Zbilut J.P. A model of biological neuron with terminal chaos and quantum like features. Chaos, Solitons and Fractals. 2006; 30: 774-780, and references therein

Conte E, Todarello O, Federici A, Vitiello F, Lopane M, Khrennikov AY, Zbilut JP. Some Remarks on an Experiment Suggesting Quantum Like Behaviour of Cognitive Entities and Formulation of an Abstract Quantum Mechanical Formalism to Describe Cognitive Entity and Its Dynamics. Chaos, Solitons and Fractals 2007; 31: 1076-1088.

Conte E. Testing Quantum Consciousness. Neuroquantology 2008; 6 (2): 126-139.

Conte E, Khrennikov AY, Todarello O, Federici A, Zbilut JP. A Preliminary Experimental Verification On the Possibility of Bell Inequality Violation in Mental States Neuroquantology 2008; 6 (3): 214-221.

Conte E., Todarello O., Federici A., Zbilut J.P. Mind States follow Quantum Mechanics during Perception and Cognition of Ambiguous Figures: a Final Experimental Confirmation.2008; arXiv:0802.1835

Conte E., Khrennikov A.Y.,.Todarello O, De Robertis R., Federici A. and Zbilut J.P. On the possibility that we think in a quantum mechanical manner: an experimental verification of existing quantum interference effects in cognitive anomaly of conjunction fallacy. Chaos and Complexity Letters. 2008; 4 (3): 217-239.

Conte E., Khrennikov A.Y., Todarello O., Federici A., Zbilut J.P. Mental States Follow Quantum Mechanics during Perception and Cognition of Ambiguous Figures. Open Systems & Information Dynamics. 2009; 16 (1): 1–17; available on line PhilPapers.

Conte E., Khrennikov A.Y., Todarello O., Federici A., Zbilut J.P. On the Existence of Quantum Wave Function and Quantum Interference Effects in Mental States: An Experimental Confirmation during Perception and Cognition in Humans. NeuroQuantology. 2009; First issue 2009 – available online.

Conte E., Khrennikov A.Y., Todarello O., Federici A., Zbilut J.P. On the Existence of Quantum Wave Function and Quantum Interference Effects in Mental States. An Experimental Confirmation during Perception and Cognition in Humans. Neuroquantology 2009; 7 (2): 204-212.

Conte E. A Reformulation of von Neumann's Postulates on Quantum Measurement by Using Two Theorems in Clifford Algebra. International Journal of Theoretical Physics 2010; DOI: 10.1007/s10773-009-0239-z

Conte E., Todarello O., Laterza V., Khrennikov A.Y., Mendolicchio L., Federici A. A preliminary experimental verification of violation of Bell inequality in a quantum model of Jung theory of personality formulated by Clifford algebra. Submitted to Journal of Consciousness exploration and research.
Conte E. On the Logical Origins of Quantum Mechanics. Neuroquantology 2011; 9 (2)
Conte E. On the Logical Origins of Quantum Mechanics Demonstrated By Using Clifford Algebra: A Proof that Quantum Interference Arises in a Clifford Algebraic Formulation of Quantum Mechanics. Electronic Journal of Theoretical Physics 2011; 8 (25): 109–126
Conte E. An Investigation on the Basic Conceptual Foundations of Quantum Mechanics by Using the Clifford Algebra. Adv. Studies Theor. Phys. 2011; 5 (11): 485 – 544
Khrennikov A.Y. Classical and quantum mechanics on information spaces with applications to cognitive, psychological, social and anomalous phenomena. Foundations of Physics 1999; 29 (7): 1065-1098  arXiv:quant-ph/0003016 [pdf, ps, other]
Bulinski A., Khrennikov A.Y. Nonclassical Total Probability Formula and Quantum Interference of Probabilities Statistics Probability Letters 2002; 70 (1): 49-58  arXiv:quant-ph/0206030 [pdf, ps, other]
Khrennikov A.Y. Probabilistic pathway representation of cognitive information Journal Theor. Biology 2004; 231: 597-613.
Khrennikov A.Y., Loubenets E. On relations between probabilities under quantum and classical measurements Foundations of Physics 2004; 34 (4): 689-704  arXiv:quant-ph/0204001 [pdf]
Khrennikov A.Y. Växjö interpretation-2003: realism of contexts arXiv:quant-ph/0401072 [pdf, ps, other]
Grib A.A., Khrennikov A.Y., Starkov K. Probability amplitude in quantum like games Journal Phys. A.: Math. Gen.2006; 39: 8461-8475 arXiv:quant-ph/0308074 [pdf, ps, other]
Khrennikov A.Y. Quantum probabilities' as context depending probabilities arXiv:quant-ph/0106073  [pdf, ps, other]
Khrennikov A.Y. Ubiquitous Quantum Structure, Springer, Verlag 2010
Kochen S., Specker E. J. Math. Mechanics 1967; 17: 59-87.
Mermin N.D. Quantum mysteries revisited. Am. J. Phys. 1990; 58: 731-734
Peres A. Two simple proofs of the Kochen-Specker theorem. J. Phys. A: Math. Gen. 1991; 24: L175-L178, and Quantum Theory: concepts and methods, Kluwer Academic Press, New York, 2002.
 Von Neumann J. Mathematical foundations of quantum mechanics, Princeton University Press, 1996

Further reading:
Landsberg P.T. Does quantum mechanics exclude life? Nature1964; 203: 928-30.